\documentclass[conference,a4paper]{IEEEtran}
\IEEEoverridecommandlockouts
\usepackage{amsmath,amsfonts,mathtools}
\usepackage[noend]{algorithmic}
\usepackage{algorithm}
\usepackage{array}
\usepackage[caption=false,font=footnotesize]{subfig}
\usepackage{textcomp}
\usepackage{stfloats}
\usepackage{url}
\usepackage{color}
\usepackage{verbatim}
\usepackage{graphicx}
\usepackage{cite}
\usepackage[hidelinks]{hyperref}
\usepackage[nolist]{acronym}
\usepackage{bbm}
\usepackage[caption=false,font=footnotesize]{subfig}
\def\BibTeX{{\rm B\kern-.05em{\sc i\kern-.025em b}\kern-.08em
    T\kern-.1667em\lower.7ex\hbox{E}\kern-.125emX}}

\makeatletter
\newcommand\fs@betterruled{%
  \def\@fs@cfont{\bfseries}\let\@fs@capt\floatc@ruled
  \def\@fs@pre{\vspace*{5pt}\hrule height.8pt depth0pt \kern2pt}%
  \def\@fs@post{\kern2pt\hrule\relax}%
  \def\@fs@mid{\kern2pt\hrule\kern2pt}%
  \let\@fs@iftopcapt\iftrue}
\floatstyle{betterruled}
\restylefloat{algorithm}
\makeatother

\graphicspath{{figures/}}

\newcommand\plotwidth{0.48}

\begin{acronym}
\acro{5G}{fifth generation}
\acro{6G}{sixth generation}
\acro{AMP}{approximate message passing}
\acro{AP}{access point}
\acro{APP}{a posteriori probability}
\acro{AWGN}{additive white Gaussian noise}
\acro{B5G}{beyond fifth generation}
\acro{BG}{Bernoulli-Gaussian}
\acro{BPSK}{binary phase-shift-keying}
\acro{CDF}{cumulative distribution function}
\acro{CF-MaMIMO}{cell-free massive multiple-input multiple-output}
\acro{CPU}{central processing unit}
\acro{CSI}{channel state information}
\acro{DER}{detection error rate}
\acro{DFT}{discrete Fourier transform}
\acro{EP}{expectation propagation}
\acro{GaBP}{Gaussian belief propagation}
\acro{GF-CF-MaMIMO}{grant-free cell-free massive multiple-input multiple-output}
\acro{iid}[i.i.d.]{independent and identically distributed}
\acro{IoT}{Internet-of-Things}
\acro{JACD}{joint activity detection, channel estimation, and data detection}
\acro{JACD-EP}{joint activity detection, channel estimation, and data detection via expectation propagation}
\acro{JAC}{joint activity detection and channel estimation}
\acro{JAC-EP}{joint activity detection and channel estimation via expectation propagation}
\acro{JACD-EP-BG}{joint activity detection, channel estimation, and data detection via expectation propagation with Bernoulli-Gaussian distributions}
\acro{JCD}{joint channel estimation and data detection}
\acro{KL}{Kullback-Leibler}
\acro{LSFC}{large-scale fading coefficient}
\acro{MaMIMO}{massive multiple-input multiple-output}
\acro{MIMO}{multiple-input multiple-output}
\acro{MAP}{maximum a posteriori}
\acro{MMSE}{minimum mean squared error}
\acro{MMV-AMP}{multiple measurement vector approximate message passing}
\acro{NMSE}{normalized mean squared error}
\acro{PC}{pilot contamination}
\acro{QAM}{quadrature amplitude modulation}
\acro{QoS}{quality of service}
\acro{SER}{symbol error rate}
\acro{UE}{user equipment}
\acro{UPP}{uniform point process}
\acro{VB}{variational Bayes}
\acro{VB-BP-EP}[VB-EP]{variational Bayes, belief propagation, and expectation propagation}
\acro{VPS}{virtual pilot sequence}
\acro{VPM}{virtual pilot matrix}
\end{acronym}

\newcommand{\lvec}[1]{\ensuremath{\mathbf{#1}}}  		% latin symbols
\newcommand{\gvec}[1]{\ensuremath{\boldsymbol{#1}}}		% greek symbols
\newcommand{\lmat}[1]{\ensuremath{\mathbf{#1}}}  		% latin symbols
\newcommand{\gmat}[1]{\ensuremath{\boldsymbol{#1}}}		% greek symbols
\newcommand{\transymb}{\ensuremath{\mathsf{T}}}
\newcommand{\hermsymb}{\ensuremath{\mathsf{H}}}
\newcommand{\tran}{\ensuremath{^{\mkern-1mu\transymb}}}
\newcommand{\herm}{\ensuremath{^{\mkern-1mu\hermsymb}}}

\newcommand{\trace}[1]{\ensuremath{\mathrm{tr}\!\left\{#1\right\}}}
\newcommand*{\argmax}{\ensuremath{\mathop{\mathrm{arg\,max}}}}
\newcommand{\est}[2]{\ensuremath{E_{#1}\!\left\{#2\right\}}}

\newcommand{\Cset}{\mathbb{C}}

\newcommand{\CN}[1]{\ensuremath{\mathcal{CN}\!\left(#1\right)}}

\newcommand{\cat}[1]{\ensuremath{\pi\!\left(#1\right)}}
\newcommand{\diag}[1]{\ensuremath{\mathrm{diag}\!\left(#1\right)}}
\newcommand{\blkdiag}[1]{\ensuremath{\mathrm{blkdiag}\!\left(#1\right)}}
\newcommand{\vect}[1]{\ensuremath{\mathrm{vec}\!\left\{#1\right\}}}
\newcommand{\ind}[1]{\ensuremath{\mathbbm{1}_{#1}}}
\newcommand{\pilotsymb}{\ensuremath{p}}
\newcommand{\datasymb}{\ensuremath{d}}
\newcommand{\pilot}[1]{\ensuremath{#1^\pilotsymb}}
\newcommand{\data}[1]{\ensuremath{#1^\datasymb}}
\newcommand{\msg}[2]{\ensuremath{m_{#1;#2}}}

\newcommand{\catmsg}[2]{\ensuremath{\pi_{#1;#2}}}
\newcommand{\Mumsg}[2]{\ensuremath{\gvec{\mu}_{#1;#2}}}
\newcommand{\Cmsg}[2]{\ensuremath{\lvec{C}_{#1;#2}}}
\newcommand{\Gammamsg}[2]{\ensuremath{\gvec{\gamma}_{#1;#2}}}
\newcommand{\Lambdamsg}[2]{\ensuremath{\gvec{\Lambda}_{#1;#2}}}
\newcommand{\Mumsga}[2]{\ensuremath{\check{\gvec{\mu}}_{#1#2}}}
\newcommand{\Cmsga}[2]{\ensuremath{\check{\lvec{C}}_{#1#2}}}
\newcommand{\Gammamsga}[2]{\ensuremath{\check{\gvec{\gamma}}_{#1#2}}}
\newcommand{\Lambdamsga}[2]{\ensuremath{\check{\gvec{\Lambda}}_{#1#2}}}
\newcommand{\Mumsgb}[2]{\ensuremath{\hat{\gvec{\mu}}_{#1#2}}}
\newcommand{\Cmsgb}[2]{\ensuremath{\hat{\lvec{C}}_{#1#2}}}
\newcommand{\Gammamsgb}[2]{\ensuremath{\hat{\gvec{\gamma}}_{#1#2}}}
\newcommand{\Lambdamsgb}[2]{\ensuremath{\hat{\gvec{\Lambda}}_{#1#2}}}
\begin{document}
\linespread{0.92}

\title{Bayesian Learning for Pilot Decontamination in Cell-Free Massive MIMO
\thanks{This work was funded by the DFG (German Research Foundation) for FAU and by the ANR for EURECOM – Project CellFree6G CO 1311/1-1, Project ID 491320625.}}

\author{\IEEEauthorblockN{Christian Forsch\IEEEauthorrefmark{1}, Zilu Zhao\IEEEauthorrefmark{2}, Dirk Slock\IEEEauthorrefmark{2}, and Laura Cottatellucci\IEEEauthorrefmark{1}}
\IEEEauthorblockA{\IEEEauthorrefmark{1}Institute for Digital Communications, Friedrich-Alexander-Universität Erlangen-Nürnberg, Erlangen, Germany \\
\IEEEauthorrefmark{2}Communication Systems Department, EURECOM, Sophia Antipolis, France \\
Email: \{christian.forsch, laura.cottatellucci\}@fau.de, \{zilu.zhao, dirk.slock\}@eurecom.fr}
}

\maketitle

\begin{abstract}
\Ac{PC} arises when the pilot sequences assigned to \acp{UE} are not mutually orthogonal, eventually due to their reuse.
In this work, we propose a novel \ac{EP}-based \ac{JCD} algorithm specifically designed to mitigate the effects of \ac{PC} in the uplink of \ac{CF-MaMIMO} systems.
This modified bilinear-\ac{EP} algorithm is distributed, scalable, demonstrates strong robustness to \ac{PC}, and outperforms state-of-the-art Bayesian learning algorithms.
Through a comprehensive performance evaluation, we assess the performance of Bayesian learning algorithms for different pilot sequences and observe that the use of non-orthogonal pilots can lead to better performance compared to shared orthogonal sequences.
Motivated by this analysis, we introduce a new metric to quantify \ac{PC} at the \ac{UE} level.
We show that the performance of the considered algorithms degrades monotonically with respect to this metric, providing a valuable theoretical and practical tool for understanding and managing \ac{PC} via iterative \ac{JCD} algorithms.
\end{abstract}


\acresetall

\section{Introduction}\label{sec:intro}
Distributed \ac{MIMO} communications, particularly in the form of cell-free massive \ac{MIMO} (\acs{CF-MaMIMO})\acused{CF-MaMIMO}\acused{MaMIMO} networks, are expected to play a key role in advancing next-generation mobile communication systems by enabling high-rate and energy-efficient communication everywhere in the coverage area~\cite{Ngo2017,Ngo2018,Mohammadi2024}.
In this network architecture, a large number of spatially distributed \acp{AP} are communicating with a smaller number of \acp{UE}.
One major challenge in practical \ac{CF-MaMIMO} networks is \ac{PC}, which arises from the use of non-orthogonal pilot sequences for channel estimation and deteriorates the overall system performance.
In real-world networks, the use of non-orthogonal pilot sequences is necessary due to the potentially large number of \acp{UE} in the network.
Ensuring orthogonality would require prohibitively long pilot sequences, reducing spectral efficiency and throughput per user.
Furthermore, unlike centralized \ac{MaMIMO}, channel hardening and favorable propagation typically do not hold in \ac{CF-MaMIMO}~\cite{Yin2014,Chen2018,Gholami2020a,Gholami2020b}, which precludes the use of existing pilot decontamination methods proposed for centralized \ac{MaMIMO}, e.g.,~\cite{Ngo2012,Yin2013,Cottatellucci2013,Mueller2014,Yin2016}.
Even though \ac{PC} has been shown not to represent a fundamental limitation for centralized and \ac{CF-MaMIMO} systems~\cite{Bjoernson2018,Polegre2021}, it still remains a significant practical challenge, especially in scalable \ac{CF-MaMIMO} systems with a limited number of \acp{AP} and \ac{AP} antennas.
This motivates the development of new effective, efficient, and distributed pilot decontamination schemes capable of leveraging other degrees of freedom beyond those considered in~\cite{Bjoernson2018} and~\cite{Polegre2021}, which rely primarily on pilot symbols and channel statistics.
These additional methods are presented below.

An efficient approach to mitigate \ac{PC} is through optimized pilot assignment schemes because strong \ac{PC} arises when \acp{UE} in close proximity utilize the same pilot sequence.
In~\cite{Ngo2017}, the authors propose a greedy pilot assignment scheme based on minimizing \ac{PC} for the user with the lowest rate.
Here, the \ac{PC} for a given \ac{UE} is quantified by the average interference power after projecting the received pilot signal onto the pilot sequence used by the given \ac{UE}.
In~\cite{Chen2021}, a K-means and a user-group-based pilot assignment scheme are presented where the distances between all \acp{UE} and \acp{AP} and their serving relationships are utilized to assign the pilot sequences.
An alternative strategy to alleviate \ac{PC} leverages \ac{JCD}.
The authors in~\cite{Song2022} developed a \ac{JCD} scheme based on forward backward splitting which exploits the sparsity of \ac{CF-MaMIMO} channels and employs non-orthogonal pilot sequences.
Bayesian learning methods have also been explored in the literature for \ac{JCD}.
In~\cite{Karataev2024}, a distributed semi-blind \ac{JCD} message-passing algorithm has been presented for \ac{CF-MaMIMO} networks which applies \ac{EP} on a factor graph.
A similar approach for grant-free \ac{CF-MaMIMO} has been presented in~\cite{Forsch2024}.
The authors in~\cite{Zhao2024} proposed a semi-blind \ac{JCD} algorithm based on Bethe free energy optimization which combines \ac{VB} and \ac{EP} referred shortly to as \acs{VB-BP-EP}\acused{VB-BP-EP} throughout this paper.

In this work, we propose a novel \ac{JCD} algorithm based on \ac{EP}, specifically designed to enhance robustness against \ac{PC} in \ac{CF-MaMIMO} systems.
The algorithm builds upon the bilinear-EP algorithm presented in\cite{Karataev2024} and incorporates a modified scheduling and message passing for bilinear structure to effectively exploit the inherent structure of the received data signals and suppress the impact of \ac{PC}.
We evaluate the proposed method against state-of-the-art Bayesian learning algorithms and demonstrate superior detection and estimation performance, particularly under severe \ac{PC}.
This thorough analysis of \ac{PC} was not conducted in prior work.
Our analysis further considers both contamination caused by non-orthogonal sequences and by the reuse of identical orthogonal sequences, revealing that Bayesian methods exhibit greater robustness when non-orthogonal pilots with low correlation are used.
This insight motivates the introduction of a novel metric tailored to quantify \ac{PC} and assess its impact on iterative \ac{JCD} algorithms.

\textit{Notation:}
$(\cdot)\tran$ and $(\cdot)\herm$ are the transpose and the conjugate transpose operator, respectively.
$\delta(\cdot)$ and $\ind{(\cdot)}$ denote the Dirac delta and the indicator function, respectively.
$\CN{\lvec{x}|\gvec{\mu},\lmat{C}}$ represents the circularly-symmetric multivariate complex Gaussian distribution of a complex-valued vector $\lvec{x}$ with mean $\gvec{\mu}$ and covariance matrix $\lmat{C}$.
$\cat{x}$ denotes the categorical distribution of a discrete random variable $x$.
The notation $x\sim p$ indicates that the random variable $x$ follows the distribution $p$.
The message sent from the factor node $\Psi_\alpha$ to the variable node $\lvec{x}_\beta$ in a factor graph is denoted as $\msg{\Psi_\alpha}{\lvec{x}_\beta}$ and consists of parameters of a distribution in the exponential family which are denoted with the same subscript of the message, e.g., mean $\Mumsg{\Psi_\alpha}{\lvec{x}_\beta}$ and covariance matrix $\Cmsg{\Psi_\alpha}{\lvec{x}_\beta}$ for a Gaussian distribution or probability values $\catmsg{\Psi_\alpha}{\lvec{x}_\beta}$ for a categorical distribution.
The same holds for variable-to-factor messages $\msg{\lvec{x}_\beta}{\Psi_\alpha}$.

\section{System Model}\label{sec:sys_mod}
We consider the uplink of a \ac{CF-MaMIMO} network consisting of $L$ geographically distributed $N$-antenna \acp{AP} and $K$ synchronized single-antenna \acp{UE}.
All \acp{AP} are connected to a \ac{CPU} via fronthaul links to share information.
During the channel coherence time of $T$ channel uses, the received signal $\lmat{Y}\!\in\!\Cset^{LN\times T}$ at all the \acp{AP} is given by
\begin{equation}
    \lmat{Y} = \lmat{H}\lmat{X} + \lmat{N}
    \label{eq:Y}
\end{equation}
where $\lmat{H}=[\lmat{H}_1\tran\cdots\lmat{H}_L\tran]\tran\!\in\!\Cset^{LN\times K}$ is the channel matrix and $\lmat{H}_l=[\lvec{h}_{l,1}\cdots\lvec{h}_{l,K}]\!\in\!\Cset^{N\times K}$ denotes the channel between \ac{AP} $l$ and all \acp{UE};
$\lmat{X}=[\lvec{x}_1\cdots\lvec{x}_K]\tran\!\in\!\Cset^{K\times T}$ is the transmit symbol matrix and $\lvec{x}_k\!\in\!\Cset^{T\times1}$ represents the transmit sequence of \ac{UE} $k$;
and $\lmat{N}\!\in\!\Cset^{LN\times T}$ is the matrix of \ac{AWGN} with \ac{iid} elements $n\sim\CN{n|0,\sigma_n^2}$.
The channels are assumed to be constant during the channel coherence time.
We assume block Rayleigh fading channels between \ac{UE} $k$ and \ac{AP} $l$, i.e., $\lvec{h}_{l,k}\sim p_{h_{l,k}}(\lvec{h}_{l,k})=\CN{\lvec{h}_{l,k}|\lvec{0},\gmat{\Xi}_{l,k}}$ where $\gmat{\Xi}_{l,k}$ is the channel correlation matrix with the \ac{LSFC} $\xi_{l,k}=\frac{1}{N}\trace{\gmat{\Xi}_{l,k}}$.
The transmit symbol matrix consists of a pilot part $\pilot{\lmat{X}}\!\in\!\Cset^{K\times T_p}$ and a data part $\data{\lmat{X}}\!\in\!\mathcal{X}^{K\times T_d}$, i.e., $\lmat{X}=[\pilot{\lmat{X}}\;\data{\lmat{X}}],$ with $T_p+T_d=T$ and $\mathcal{X}$ being the transmit data symbol constellation of cardinality $M=|\mathcal{X}|$.
The average transmit symbol power is given by $\sigma_x^2=\est{}{|x_{kt}|^2}$.
A similar decomposition in pilot and data part applies to the receive matrix, i.e., $\lmat{Y}=[\pilot{\lmat{Y}}\;\data{\lmat{Y}}]$ with received pilots $\pilot{\lmat{Y}}\!\in\!\Cset^{L\times T_p}$ and received data $\data{\lmat{Y}}\!\in\!\Cset^{L\times T_d}$.
Furthermore, we assume that the pilot length is smaller than the number of \acp{UE}, $T_p<K$, since in practice the number of \acp{UE} can be very large and, thus, it is not practical to assign orthogonal pilot sequences to the \acp{UE}.
This gives rise to the \ac{PC} effect.

\section{Problem Formulation}\label{subsec:problem}
Due to PC, channel estimation based solely on pilot sequences typically presents severely degraded performance. An effective strategy to mitigate the impact of PC is to leverage not only the pilot sequences but also the detected data symbols to iteratively refine both channel estimation and data detection. In this context, the receiver's task is to jointly estimate the channel matrix $\lmat{H}$ and the user data matrix $\data{\lmat{X}}$.
The \ac{MAP} estimator is given by
\begin{equation}
    (\hat{\lmat{H}},\data{\hat{\lmat{X}}}) = \argmax_{\lmat{H},\data{\lmat{X}}}\;p_\text{APP}(\lmat{H},\data{\lmat{X}}),
    \label{eq:MAP}
\end{equation}
with the \ac{APP} distribution $p_\text{APP}(\lmat{H},\data{\lmat{X}})$ which can be factorized by applying Bayes' theorem,
\begin{align}
    p_\text{APP}(\lmat{H},\data{\lmat{X}})&=p_{H,\data{X}|Y,\pilot{X}}(\lmat{H},\data{\lmat{X}}|\lmat{Y},\pilot{\lmat{X}})\nonumber\\
    &\propto p_{Y|H,X}(\lmat{Y}|\lmat{H},\lmat{X})\cdot p_{H}(\lmat{H})\cdot p_{X}(\lmat{X}).
    \label{eq:APP}
\end{align}
Solving the inference problem in~\eqref{eq:MAP} is computationally intractable due to the high dimensionality of the involved variables.
To address this challenge, we employ Bayesian learning techniques to obtain tractable approximations of the \ac{MAP} estimates.

\section{Bilinear-EP Algorithm}\label{sec:bilinear-EP}
In this section, we propose a novel \ac{JCD} algorithm with enhanced performance under \ac{PC}. It builds upon the bilinear-\ac{EP} algorithm introduced in~\cite{Karataev2024} for distributed semi-blind \ac{JCD} in \ac{CF-MaMIMO} systems. Unlike the baseline, our approach incorporates knowledge and observations of the pilot sequences in the message-passing procedure.
In the following, we present the underlying factorization and message-passing steps, and refer to~\cite{Karataev2024} for general details on bilinear-EP and message derivations. 

\subsection{Factor Graph Representation}\label{subsec:FG}
\begin{figure}[t]
    \centerline{\includegraphics[width=0.4\textwidth]{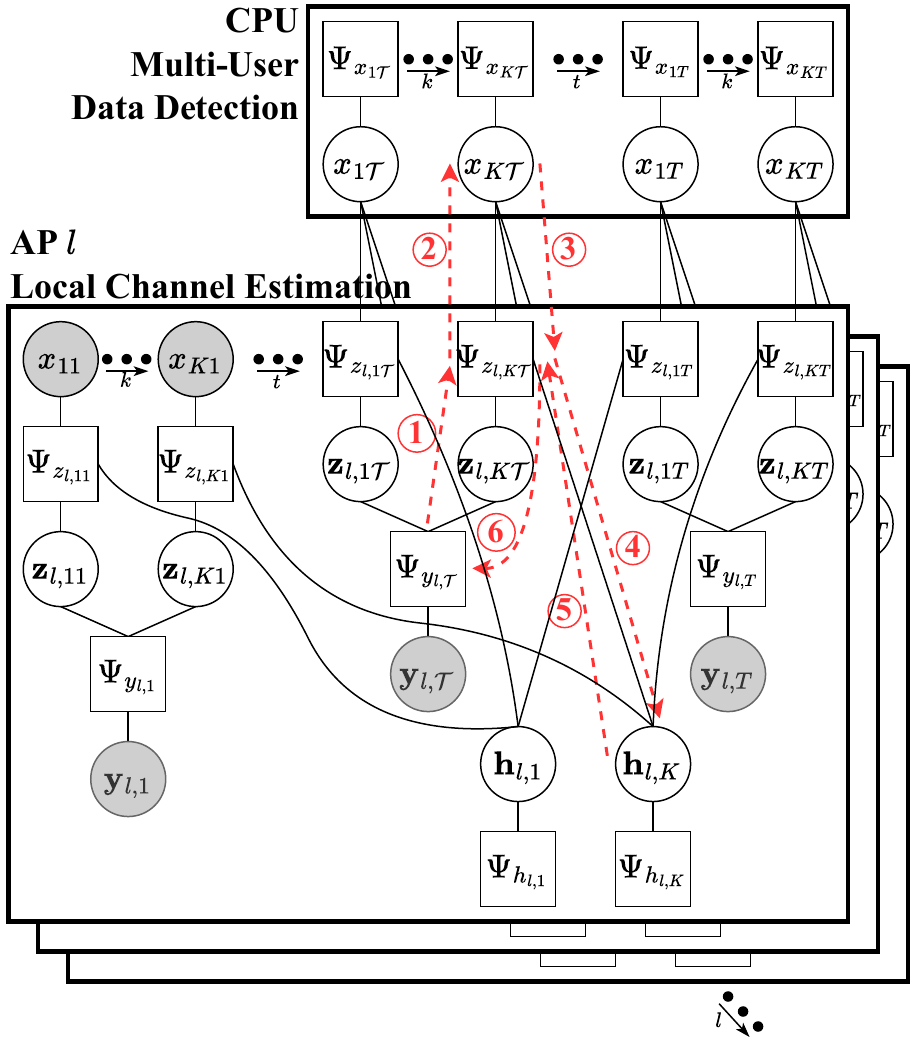}}
    \caption{Factor graph for bilinear-\ac{EP} with $\mathcal{T}\coloneq T_p+1$. The numbered red dashed arrows show the message update scheduling according to Algorithm~\ref{alg:bilinear-EP}.}
    \vspace*{-1mm}
    \label{fig:FG}
\end{figure}

In order to solve the \ac{MAP} problem in~\eqref{eq:MAP}, the auxiliary variables $\lvec{z}_{l,kt} \coloneq \lvec{h}_{l,k}x_{kt}$ $\forall l,k,t$ are introduced and stored in the array $\lmat{Z}$.
The \ac{APP} distribution with respect to the channel, data, and auxiliary variables can be factorized as follows,
\begin{equation}
\begin{split}
    &p_\text{APP}(\lmat{H},\data{\lmat{X}},\lmat{Z})\propto\prod_{l=1}^L\prod_{k=1}^K\prod_{t=1}^T\Big[p(\lvec{y}_{l,t}|\lvec{z}_{l,1t},...,\lvec{z}_{l,Kt})\\
    &\qquad\qquad\quad\cdot p(\lvec{z}_{l,kt}|\lvec{h}_{l,k},x_{kt})\cdot\tilde{p}_{h_{l,k}}(\lvec{h}_{l,k})\cdot p_x(x_{kt})\Big],
    \label{eq:APP_aux}
\end{split}
\end{equation}
where the independence of channel vectors for different \acp{AP} and \acp{UE} as well as the independence of data symbols for different \acp{UE} and time indices is exploited.
The probability distribution $\tilde{p}_{h_{l,k}}(\lvec{h}_{l,k})$ represents modified channel information compared to the prior information $p_{h_{l,k}}(\lvec{h}_{l,k})$.
How to obtain this modified information will be explained in the following.
The factor graph representing the \ac{APP} distribution~\eqref{eq:APP_aux} is shown in Fig.~\ref{fig:FG}. It comprises variable nodes (circles) and factor nodes (rectangles), organized according to their implementation at the CPU and the \acp{AP}. Each factor node corresponds to one of the following probability distributions,
\begin{alignat*}{2}
    &\Psi_{y_{l,t}} &&\coloneq p(\lvec{y}_{l,t}|\lvec{z}_{l,1t},...,\lvec{z}_{l,Kt}) =
    \CN{\!\lvec{y}_{l,t}\Big|\sum_{k=1}^K\lvec{z}_{l,kt},\sigma_n^2\lmat{I}_N\!}\!,
    \\
    &\Psi_{z_{l,kt}} &&\coloneq p(\lvec{z}_{l,kt}|\lvec{h}_{l,k},x_{kt}) = \delta(\lvec{z}_{l,kt}-\lvec{h}_{l,k}x_{kt}),
    \\
    &\Psi_{h_{l,k}} &&\coloneq \tilde{p}_{h_{l,k}}(\lvec{h}_{l,k}) = \CN{\lvec{h}_{l,k}\big|\tilde{\gvec{\mu}}_{h_{l,k}},\tilde{\lmat{C}}_{h_{l,k}}}\!,
    \\
    &\Psi_{x_{kt}} &&\coloneq p_x(x_{kt}) =
    \begin{cases}
 {\ind{x_{kt}=\pilot{x}_{kt}}}&\text{for }t\leq T_p\\
 \frac{1}{M}\ind{x_{kt}\in\mathcal{X}}&\text{for }t>T_p
    \end{cases}.
\end{alignat*}
Here, $\tilde{p}_{h_{l,k}}(\lvec{h}_{l,k})$ represents the modified prior channel distribution obtained solely from processing the pilot sequences.
In this work, we employ the \ac{MMSE} estimator which can be applied independently for each \ac{AP} without loss of optimality.
The resulting estimates are denoted by $\tilde{\gvec{\mu}}_{h_{l,k}}$ with corresponding error covariance matrix $\tilde{\lmat{C}}_{h_{l,k}}$ given by~\cite{Kay1993}
\begin{align}
    \tilde{\gvec{\mu}}_{h_{l,k}} &= \Big[\lmat{F}\vect{\pilot{\lmat{Y}}_l}\Big]_{\lvec{i}_k,1},
    \label{eq:mu_MMSE}\\
    \tilde{\lmat{C}}_{h_{l,k}} &= \left[\left(\lmat{I}_{KN}-\lmat{F}\pilot{\tilde{\lmat{X}}}\right)\gmat{\Xi}_l\right]_{\lvec{i}_k,\lvec{i}_k},
    \label{eq:C_MMSE}
\end{align}
where $\lmat{F}=\gmat{\Xi}_l\tilde{\lmat{X}}^{\pilotsymb\:\!\hermsymb}(\pilot{\tilde{\lmat{X}}}\gmat{\Xi}_l\tilde{\lmat{X}}^{\pilotsymb\:\!\hermsymb}+\sigma_n^2\lmat{I}_{NT_p})^{-1}$, $\pilot{\tilde{\lmat{X}}}={\pilot{\lmat{X}}}\tran\otimes\lmat{I}_N$, and $\gmat{\Xi}_l=\blkdiag{\gmat{\Xi}_{l,1},\dots,\gmat{\Xi}_{l,K}}$.
The operator $\vect{\cdot}$ denotes vectorization,
$[\cdot]_{\lvec{i},\lvec{j}}$ extracts elements located at rows and columns specified by the indices $\lvec{i}$ and $\lvec{j}$, respectively, of the matrix in the square brackets, $\lvec{i}_k=(k-1)N+1:kN$ represents the index range corresponding to \ac{UE} $k$, and the function $\blkdiag{\cdot}$ constructs a block diagonal matrix from its arguments.

The key modification of the proposed approach compared to the bilinear-\ac{EP} algorithm in~\cite{Karataev2024} lies in the inclusion of pilot symbols within the factor graph and, hence, within the message-passing procedure.
The corresponding message update rules are detailed in the following section.

\subsection{Message Updates and Scheduling}\label{subsec:MP_updates}
The bilinear-\ac{EP} algorithm models the variables $x_{kt}$ with categorical distributions, while $\lvec{z}_{l,kt}$ and $\lvec{h}_{l,k}$ are modeled as multivariate complex Gaussian distributions. 
Accordingly, \ac{EP} message-passing update rules are applied to the factor graph in Fig.~\ref{fig:FG}.
We present the message scheduling and the final update rules. Detailed derivations can be found in~\cite{Karataev2024}.
Note that the mean vector $\gvec{\mu}$ and the covariance matrix $\lmat{C}$ of a Gaussian random variable are equivalently represented by the natural parameters $\gvec{\gamma}\!=\!\lmat{C}^{-1}\gvec{\mu}$ and $\gmat{\Lambda}\!=\!\lmat{C}^{-1}$.
In the following, we will interchangeably use both these  representations without explicitly stating the transformation, i.e., if $\Mumsg{\Psi_\alpha}{\lvec{x}_\beta}$ and $\Cmsg{\Psi_\alpha}{\lvec{x}_\beta}$ are computed, then $\Gammamsg{\Psi_\alpha}{\lvec{x}_\beta}$ and $\Lambdamsg{\Psi_\alpha}{\lvec{x}_\beta}$ are automatically given and vice versa.

The message initialization is performed as follows: parameters describing the messages $\!\msg{\Psi_{h_{l,k}}}{\lvec{h}_{l,k}}\!$, $\!\msg{\lvec{h}_{l,k}}{\Psi_{z_{l,kt}}}\!$, and $\!\msg{\Psi_{z_{l,kt}}}{\lvec{z}_{l,kt}}\!$ $\forall k,l,t$ are initialized per Algorithm~\ref{alg:bilinear-EP}.
All other messages are initialized in an uninformative way.

The messages $\msg{\Psi_{y_{l,t}}}{\lvec{z}_{l,kt}}$ $\forall k,l,t$ are updated first, in which each \ac{AP} performs interference cancellation on the received signal using the current knowledge of the auxiliary variables $\lvec{z}_{l,kt}$,
\begin{align}
    \Mumsg{\Psi_{y_{l,t}}}{\lvec{z}_{l,kt}} &= \lvec{y}_{l,t}-\sum_{k'\neq k}\Mumsg{\Psi_{z_{l,k't}}}{\lvec{z}_{l,k't}},
    \label{eq:mu_Psi_y_z}\\
    \Cmsg{\Psi_{y_{l,t}}}{\lvec{z}_{l,kt}} &= \sigma_n^2\lmat{I}_N+\sum_{k'\neq k}\Cmsg{\Psi_{z_{l,k't}}}{\lvec{z}_{l,k't}}.
    \label{eq:C_Psi_y_z}
\end{align}

The updated information on the variables $\lvec{z}_{l,kt}$ at each  \ac{AP} $l$ is used to refine the local beliefs on the data symbols $x_{kt},$ which are subsequently shared with the \ac{CPU}.
This is done by updating the message $\msg{\Psi_{z_{l,kt}}}{x_{kt}}$ $\forall k,l,t>T_p$,
\begin{equation}
    \catmsg{\Psi_{z_{l,kt}}}{x_{kt}}(x_{kt}) \propto \theta(x_{kt}),
    \label{eq:m_Psi_z_x}
\end{equation}
with
\begin{equation*}
\begin{split}
    \theta(x_{kt}) = \mathcal{CN}(\lvec{0}|&\Mumsg{\Psi_{y_{l,t}}}{\lvec{z}_{l,kt}}-\Mumsg{\lvec{h}_{l,k}}{\Psi_{z_{l,kt}}}x_{kt},\\
    &\Cmsg{\Psi_{y_{l,t}}}{\lvec{z}_{l,kt}}+\Cmsg{\lvec{h}_{l,k}}{\Psi_{z_{l,kt}}}|x_{kt}|^2).
    \label{eq:theta_mmd_Psi_z_z}
\end{split}
\end{equation*}

Next, the messages $\msg{x_{kt}}{\Psi_{z_{l,kt}}}$ $\forall k,l,t>T_p$ are updated by aggregating the data symbol beliefs from all \acp{AP} at the \ac{CPU}, which then sends the following refined beliefs back to the \acp{AP},
\begin{equation}
    \catmsg{x_{kt}}{\Psi_{z_{l,kt}}}(x_{kt}) \propto \prod_{l'\neq l}\catmsg{\Psi_{z_{l',kt}}}{x_{kt}}(x_{kt}).
    \label{eq:m_x_Psi_z}
\end{equation}

The updated beliefs on the symbols $x_{kt}$ at each \ac{AP} are then used to refine the channel estimates $\lvec{h}_{l,k}.$ This refinement is achieved through the update of the message $\msg{\Psi_{z_{l,kt}}}{\lvec{h}_{l,k}}$ $\forall k,l,t$, given by
\begin{align}
    \Lambdamsg{\Psi_{z_{l,kt}}}{\lvec{h}_{l,k}} &= \Lambdamsgb{1_{l,kt}}{}-\Lambdamsg{\lvec{h}_{l,k}}{\Psi_{z_{l,kt}}},
    \label{eq:C_Psi_z_h}\\
    \Gammamsg{\Psi_{z_{l,kt}}}{\lvec{h}_{l,k}} &= \Gammamsgb{1_{l,kt}}{}-\Gammamsg{\lvec{h}_{l,k}}{\Psi_{z_{l,kt}}},
    \label{eq:mu_Psi_z_h}
\end{align}
with $\Mumsgb{1_{l,kt}}{}=\frac{\Mumsga{{l,kt}}{}(\pilot{x}_{kt})}{\pilot{x}_{kt}}$, $\Cmsgb{1_{l,kt}}{}=\frac{\Cmsga{{l,kt}}{}(\pilot{x}_{kt})}{|\pilot{x}_{kt}|^2}$ for $t\leq T_p$ and
\begin{align}
    \Mumsgb{1_{l,kt}}{} &= \frac{1}{{Z}_{l,kt}}\!\sum_{x_{kt}\in\mathcal{X}}\!\catmsg{x_{kt}}{\Psi_{z_{l,kt}}}(x_{kt})\cdot\frac{\theta(x_{kt})}{x_{kt}}\cdot\Mumsga{{l,kt}}{}(x_{kt}),
    \nonumber\\
\begin{split}
    \Cmsgb{1_{l,kt}}{} &= \frac{1}{{Z}_{{l,kt}}}\!\sum_{x_{kt}\in\mathcal{X}}\!\catmsg{x_{kt}}{\Psi_{z_{l,kt}}}(x_{kt})\cdot\frac{\theta(x_{kt})}{|x_{kt}|^2}\cdot\big(\Cmsga{{l,kt}}{}(x_{kt})\\
    &\qquad\qquad+\Mumsga{{l,kt}}{}(x_{kt})\cdot\Mumsga{{l,kt}}{}\herm(x_{kt})\big) - \Mumsgb{{l,kt}}{}\Mumsgb{{l,kt}}{}\herm,
    \nonumber
\end{split}
\end{align}
for $t>T_p$ with
\begin{align}
    {Z}_{{l,kt}} &= \!\sum_{x_{kt}\in\mathcal{X}}\!\catmsg{x_{kt}}{\Psi_{z_{l,kt}}}(x_{kt})\cdot\theta(x_{kt}),
    \nonumber\\
    \Lambdamsga{{l,kt}}{}(x_{kt}) &= \Lambdamsg{\Psi_{y_{l,t}}}{\lvec{z}_{l,kt}}+\Lambdamsg{\lvec{h}_{l,k}}{\Psi_{z_{l,kt}}}|x_{kt}|^{-2},
    \nonumber\\
    \Gammamsga{{l,kt}}{}(x_{kt}) &= \Gammamsg{\Psi_{y_{l,t}}}{\lvec{z}_{l,kt}}+\Gammamsg{\lvec{h}_{l,k}}{\Psi_{z_{l,kt}}}\frac{x_{kt}}{|x_{kt}|^2}.
    \nonumber
\end{align}

Then, the messages $\msg{\lvec{h}_{l,k}}{\Psi_{z_{l,kt}}}$ $\forall k,l,t$ are updated  yielding new estimates of $\lvec{h}_{l,k}$ by combining the  information acquired across all time slots with the prior channel information,
\begin{align}
    \Lambdamsg{\lvec{h}_{l,k}}{\Psi_{z_{l,kt}}} &= \Lambdamsg{\Psi_{h_{l,k}}}{\lvec{h}_{l,k}}+\sum_{t'\neq t}\Lambdamsg{\Psi_{z_{l,kt'}}}{\lvec{h}_{l,k}},
    \label{eq:C_h_Psi_z}\\
    \Gammamsg{\lvec{h}_{l,k}}{\Psi_{z_{l,kt}}} &= \Gammamsg{\Psi_{h_{l,k}}}{\lvec{h}_{l,k}}+\sum_{t'\neq t}\Gammamsg{\Psi_{z_{l,kt'}}}{\lvec{h}_{l,k}}.
    \label{eq:mu_h_Psi_z}
\end{align}

The messages $\msg{\Psi_{z_{l,kt}}}{\lvec{z}_{l,kt}}$ $\forall k,l,t$ are updated last in an \ac{EP} iteration, generating refined estimates of the variables $\lvec{z}_{l,kt}$ which are then utilized for interference cancellation in the next iteration,
\begin{align}
    \Lambdamsg{\Psi_{z_{l,kt}}}{\lvec{z}_{l,kt}} &= \Lambdamsgb{2_{l,kt}}{}-\Lambdamsg{\Psi_{y_{l,t}}}{\lvec{z}_{l,kt}},
    \label{eq:C_Psi_z_z}\\
    \Gammamsg{\Psi_{z_{l,kt}}}{\lvec{z}_{l,kt}} &= \Gammamsgb{2_{l,kt}}{}-\Gammamsg{\Psi_{y_{l,t}}}{\lvec{z}_{l,kt}},
    \label{eq:mu_Psi_z_z}
\end{align}
with $\Mumsgb{2_{l,kt}}{}\!\!=\!\Mumsga{{l,kt}}{}(\pilot{x}_{kt})$, $\Cmsgb{2_{l,kt}}{}\!\!=\!\Cmsga{{l,kt}}{}(\pilot{x}_{kt})$ for $t\leq T_p$ and
\begin{align}
    \Mumsgb{2_{l,kt}}{} &= \frac{1}{{Z}_{{l,kt}}}\!\sum_{x_{kt}\in\mathcal{X}}\!\catmsg{x_{kt}}{\Psi_{z_{l,kt}}}(x_{kt})\cdot\theta(x_{kt})\cdot\Mumsga{{l,kt}}{}(x_{kt}),
    \nonumber\\[-1mm]
\begin{split}
    \Cmsgb{2_{l,kt}}{} &= \frac{1}{{Z}_{{l,kt}}}\!\sum_{x_{kt}\in\mathcal{X}}\!\catmsg{x_{kt}}{\Psi_{z_{l,kt}}}(x_{kt})\cdot\theta(x_{kt})\cdot\big(\Cmsga{{l,kt}}{}(x_{kt})\\
    &\qquad\qquad+\Mumsga{{l,kt}}{}(x_{kt})\cdot\Mumsga{{l,kt}}{}\herm(x_{kt})\big) - \Mumsgb{1_{l,kt}}{}\Mumsgb{1_{l,kt}}{}\herm,
\end{split}
    \nonumber
\end{align}
for $t>T_p$ with ${Z}_{{l,kt}}$, $\Lambdamsga{{l,kt}}{}(x_{kt})$, and $\Gammamsga{{l,kt}}{}(x_{kt})$ as given before.

The modified bilinear-\ac{EP} algorithm is summarized in Algorithm~\ref{alg:bilinear-EP}.
\begin{algorithm}[t]
\caption{Modified Bilinear-\ac{EP} Algorithm}
\begin{algorithmic}[1]
\renewcommand{\algorithmicrequire}{\textbf{Input:}}
\renewcommand{\algorithmicensure}{\textbf{Output:}}
\REQUIRE Pilot matrix $\pilot{\lmat{X}}$, transmit power $\sigma_x^2$, received signal $\lmat{Y}$, noise variance $\sigma_n^2$, prior distribution $\tilde{p}_{h_{l,k}}(\lvec{h}_{l,k})\equiv(\tilde{\gvec{\mu}}_{h_{l,k}},\tilde{\lmat{C}}_{h_{l,k}})$.
\ENSURE Estimated channels $\hat{\lvec{h}}_{l,k}$ and data $\data{\hat{x}}_{kt}$.
\STATE $\forall k,l,t$: Initialize all messages uninformatively except\\
$\Mumsg{\Psi_{h_{l,k}}}{\lvec{h}_{l,k}} \!= \Mumsg{\lvec{h}_{l,k}}{\Psi_{z_{l,kt}}} \!= \tilde{\gvec{\mu}}_{h_{l,k}}$,\\
$\Cmsg{\Psi_{h_{l,k}}}{\lvec{h}_{l,k}} \!\!= \Cmsg{\lvec{h}_{l,k}}{\Psi_{z_{l,kt}}} \!\!= \tilde{\lmat{C}}_{h_{l,k}}$,\\
$\Mumsg{\Psi_{z_{l,kt}}}{\lvec{z}_{l,kt}} \!\!= \begin{cases}
    \tilde{\gvec{\mu}}_{h_{l,k}}\pilot{x}_{kt}&\text{for }t\leq T_p\\
    \lvec{0}&\text{for }t>T_p
\end{cases}$,\\
$\Cmsg{\Psi_{z_{l,kt}}}{\lvec{z}_{l,kt}} \!\!= \begin{cases}
    \tilde{\lmat{C}}_{h_{l,k}}|\pilot{x}_{kt}|^2&\text{for }t\leq T_p\\
    \big(\tilde{\lmat{C}}_{h_{l,k}}++\tilde{\gvec{\mu}}_{h_{l,k}}\tilde{\gvec{\mu}}_{h_{l,k}}\herm\big)\sigma_x^2&\text{for }t>T_p
\end{cases}.$
\FOR {$i = 1$ to $i_\text{max}$}
\STATE $\forall k,l,t$: Update $\msg{\Psi_{y_{l,t}}}{\lvec{z}_{l,kt}}$ via \eqref{eq:mu_Psi_y_z}, \eqref{eq:C_Psi_y_z}.
\STATE $\forall k,l,t>T_p$: Update $\msg{\Psi_{z_{l,kt}}}{x_{kt}}$ via \eqref{eq:m_Psi_z_x}.
\STATE $\forall k,l,t>T_p$: Update $\msg{x_{kt}}{\Psi_{z_{l,kt}}}$ via \eqref{eq:m_x_Psi_z}.
\STATE $\forall k,l,t$: Update $\msg{\Psi_{z_{l,kt}}}{\lvec{h}_{l,k}}$ via \eqref{eq:C_Psi_z_h}, \eqref{eq:mu_Psi_z_h}.\label{alg_line:m_Psi_z_h}
\STATE $\forall k,l,t$: Update $\msg{\lvec{h}_{l,k}}{\Psi_{z_{l,kt}}}$ via \eqref{eq:C_h_Psi_z}, \eqref{eq:mu_h_Psi_z}.
\STATE $\forall k,l,t$: Update $\msg{\Psi_{z_{l,kt}}}{\lvec{z}_{l,kt}}$ via \eqref{eq:C_Psi_z_z}, \eqref{eq:mu_Psi_z_z}.\label{alg_line:m_Psi_z_z}
\ENDFOR
\RETURN $\hat{\lvec{h}}_{l,k}$ calculated via \eqref{eq:estimate_h} $\forall k,l$.
\RETURN $\data{\hat{x}}_{kt}$ calculated via \eqref{eq:estimate_x} $\forall k,t$.
\end{algorithmic} 
\label{alg:bilinear-EP}
\end{algorithm}
Compared to the \ac{EP} algorithm presented in~\cite{Karataev2024}, it contains additional and augmented message updates.
To be more precise, the updates of the messages $\msg{\Psi_{y_{l,t}}}{\lvec{z}_{l,kt}}$, $\msg{\Psi_{z_{l,kt}}}{\lvec{h}_{l,k}}$, $\msg{\lvec{h}_{l,k}}{\Psi_{z_{l,kt}}}$, and $\msg{\Psi_{z_{l,kt}}}{\lvec{z}_{l,kt}}$ are now also considered for $t\leq T_p$ i.e., for the pilot part as well.
Furthermore, the update of the message $\msg{\lvec{h}_{l,k}}{\Psi_{z_{l,kt}}}$ is enhanced by taking into account the additional information from $t\leq T_p$.

After performing the final \ac{EP} iteration, the channel and data estimates are computed as follows,
\begin{align}
    \hat{\lvec{h}}_{l,k} &= \hat{\lmat{\Lambda}}_{\lvec{h}_{l,k}}^{-1}\hat{\lvec{\gamma}}_{\lvec{h}_{l,k}},
    \label{eq:estimate_h}\\
    \data{\hat{x}}_{kt} &= \argmax_{\data{x}_{kt}\!\in\!\mathcal{X}}\;\hat{p}_{x_{kt}}(\data{x}_{kt}),
    \label{eq:estimate_x}
\end{align}
with
\begin{align}
    \hat{\lmat{\Lambda}}_{\lvec{h}_{l,k}} &= \Lambdamsg{\Psi_{h_{l,k}}}{\lvec{h}_{l,k}}+\sum_{t=1}^T\Lambdamsg{\Psi_{z_{l,kt}}}{\lvec{h}_{l,k}},
    \label{eq:posterior_h_Lambda}\\
    \hat{\lvec{\gamma}}_{\lvec{h}_{l,k}} &= \Gammamsg{\Psi_{h_{l,k}}}{\lvec{h}_{l,k}}+\sum_{t=1}^T\Gammamsg{\Psi_{z_{l,kt}}}{\lvec{h}_{l,k}},
    \label{eq:posterior_h_gamma}
\end{align}
and the approximated posterior data distribution
\begin{align}
    \hat{p}_{x_{kt}}(\data{x}_{kt}) &\propto \prod_{l=1}^L\catmsg{\Psi_{z_{l,kt}}}{x_{kt}}(\data{x}_{kt}).
    \label{eq:posterior_x}
\end{align}

We note that damping is applied to factor-to-variable messages using a damping parameter $\eta\!\in\![0,1]$ to improve the stability of the bilinear-\ac{EP} algorithm~\cite{Karataev2024}, i.e., each updated parameter is computed as a convex combination of its previous and newly computed values.
Furthermore, the parameters of the messages $\msg{\Psi_{z_{l,kt}}}{\lvec{h}_{l,k}}$ and $\msg{\Psi_{z_{l,kt}}}{\lvec{z}_{l,kt}}$ in line~\ref{alg_line:m_Psi_z_h} and \ref{alg_line:m_Psi_z_z} of Algorithm~\ref{alg:bilinear-EP} are updated only if the corresponding covariance/precision matrices obtained by~\eqref{eq:C_Psi_z_h} and~\eqref{eq:C_Psi_z_z}, respectively, are symmetric positive definite.
Otherwise, the parameters from the previous iteration are retained.

\section{Quantifying Pilot Contamination}\label{sec:quant_PC}
In this section, we introduce a metric to quantify \ac{PC} in \ac{CF-MaMIMO} networks.
The level of \ac{PC} is influenced by the choice of the pilot matrix $\pilot{\lmat{X}}$ and the resulting correlations between pilot sequences.
Mutual coherence, which measures the similarity between pilot sequences,  was used in~\cite{Iimori2021} for pilot design to mitigate \ac{PC}.
However, due to the distributed architecture of \ac{CF-MaMIMO},  user signals can often be separated spatially, especially when the users are far apart, resulting in negligible interference.
Therefore, the spatial power profiles  of all \acp{UE}, captured by the \acp{LSFC} $\xi_{l,k}$, are also critical for characterizing \ac{PC}.
This motivates us to develop a new \ac{PC} metric which is particularly suited for \ac{JCD} in distributed systems.
Inspired by the \ac{NMSE} of the pilot-based \ac{MMSE} channel estimator, we define the \ac{PC} metric $c_k$ for \ac{UE} $k$ as
\begin{equation}
    c_k = \min_l \frac{\left[\left(\diag{\xi_{l,1},\dots,\xi_{l,K}}^{-1} + \pilot{\lmat{X}}{\pilot{\lmat{X}}}\herm\sigma_n^{-2}\right)^{-1}\right]_{k,k}}{\xi_{l,k}},
    \label{eq:c_k}
\end{equation}
where $\diag{\cdot}$ denotes a diagonal matrix with its inputs on the main diagonal.
The rationale for taking the minimum value over all \acp{AP} is that a strong, low-contamination link to any single \ac{AP} is sufficient for reliable channel estimation and successful data detection. This reliable information can be used to cancel the interference caused by the corresponding \ac{UE} and, hence, iteratively remove \ac{PC}.
This concept was formalized in~\cite{Gholami2021a} where sufficient and necessary conditions for semi-blind identifiability were established.

\section{Numerical Results}\label{sec:sims}
In this section, we present Monte Carlo simulation results for the modified bilinear-\ac{EP} algorithm and several state-of-the-art benchmark algorithms.
We consider a network spanning an area of $400\times400\,$m$^2$ comprising $L=16$ single-antenna \acp{AP}, i.e., $N=1$, placed on a regular grid at coordinates $\{(50+i\!\cdot\!100,50+j\!\cdot\!100)\,\text{m}\,|\,i,j\!\in\!\{0,1,2,3\}\}$ and placed at a height of $10\,$m.
A total of $K=8$ \acp{UE} are placed uniformly at random ground locations and transmit $T_p=4$ pilot symbols and $T_d\in\{10,30\}$ 4-\ac{QAM} uncoded data symbols.
The values of $K$ and $T_p$ are chosen such that \ac{PC} occurs and the complexity of the simulations is not too high.
In practice, longer channel coherence times allow for an increase in the number of pilot symbols $T_p$ but also for an increase in the number of \acp{UE} $K$, especially when the number of \ac{AP} antennas $LN$ is increased as well, such that \ac{PC} still occurs in these more practical cases and needs to be mitigated.
We consider two different choices of pilot sequences, referred to as Hadamard and \ac{DFT} pilots.
For Hadamard pilots, $T_p$ orthogonal Hadamard pilot sequences are considered and shared among the $K$ \acp{UE}.
For \ac{DFT} pilots, the pilot matrix $\pilot{\lmat{X}}$ is obtained by truncating a $K\times K$ DFT matrix to the first $T_p$ columns, resulting in non-orthogonal sequences.
Hence, in our simulations, the set of Hadamard pilots consists of four orthogonal pilot sequences, each of which is shared between two users.
In contrast, the set of \ac{DFT} pilots consists of eight unique but non-orthogonal pilot sequences.
The receiver noise power at each \ac{AP} is set to $\sigma_n^2=-96\,$dBm.
The \acp{LSFC} are obtained using the 3GPP urban microcell model which incorporates correlated shadow fading~\cite{Bjoernson2020}.

\begin{figure}[t]
    \centerline{\includegraphics[width=\plotwidth\textwidth]{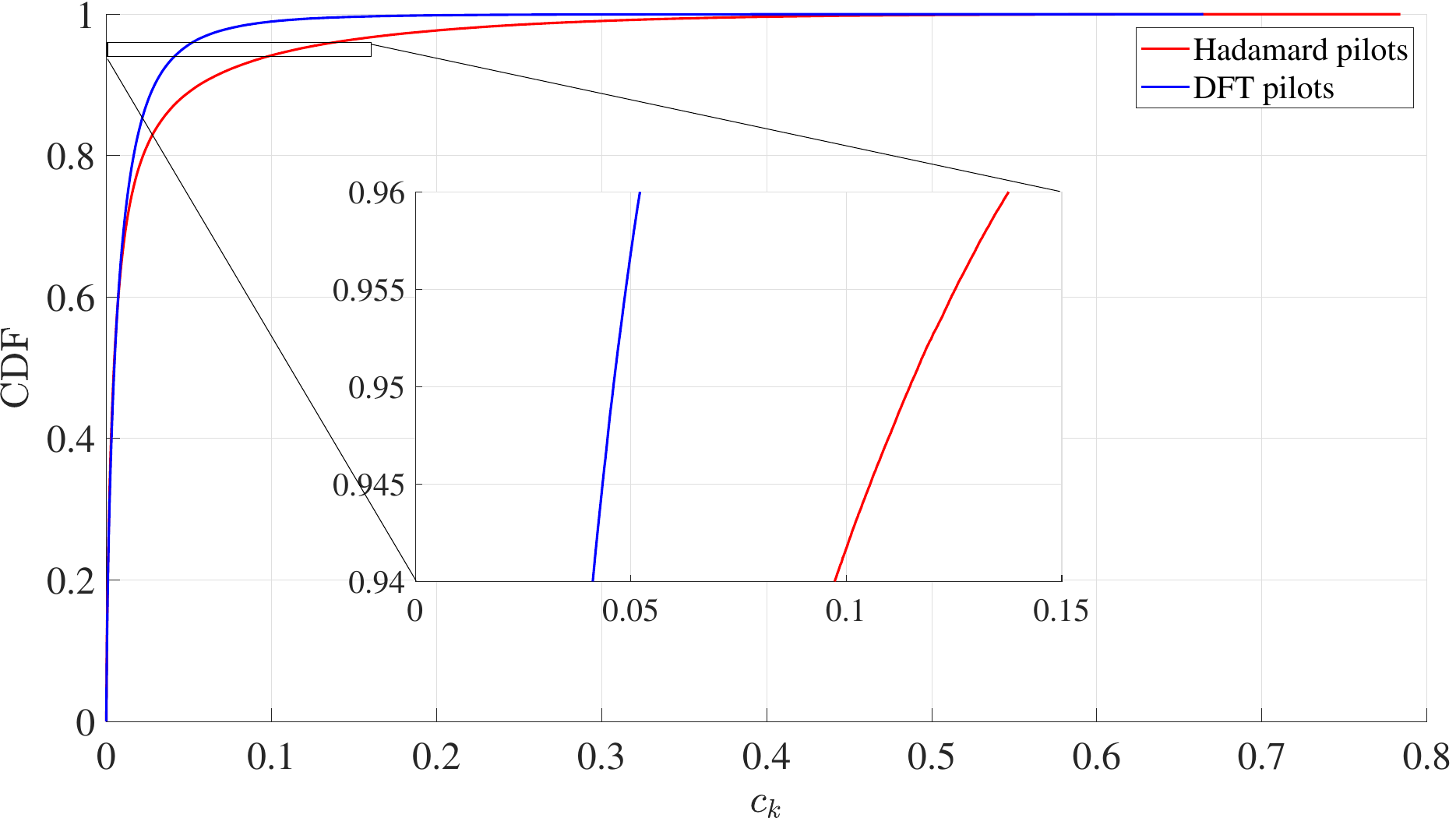}}
    \caption{\acs{CDF} of $c_k$ for different pilot sequences.}
    \vspace*{-1mm}
    \label{fig:c_k_CDF}
\end{figure}

The first set of results pertains to the \ac{PC} metric introduced in~\eqref{eq:c_k} and evaluated over $10^5$ independent large-scale fading realizations.
The \ac{CDF} of the \ac{PC} metric $c_k$ is illustrated in Fig.~\ref{fig:c_k_CDF}.
It can be observed that the non-orthogonal \ac{DFT} pilots result in lower \ac{PC} than the orthogonal Hadamard pilots which are reused among the \acp{UE}.

In the following, we show the channel estimation and data detection performance in terms of the \ac{NMSE} of the channel estimates and the \ac{SER}.
We compare the proposed modified bilinear-\ac{EP} algorithm with the bilinear-\ac{EP} algorithm in~\cite{Karataev2024}, the \ac{VB-BP-EP} algorithm presented in~\cite{Zhao2024}, and different \ac{MMSE} estimators.
For channel estimation, we consider the pilot-based \ac{MMSE} estimator and the genie-aided \ac{MMSE} estimator with perfect knowledge of the transmitted data symbols.
For symbol detection, we employ the centralized \ac{MMSE} \ac{MIMO} detector with \ac{CSI} obtained by the pilot-based \ac{MMSE} estimator and with perfect \ac{CSI}.
Note that the \ac{MMSE} detector with pilot-based \ac{CSI} serves as an upper performance bound for any linear detection scheme with imperfect \ac{CSI}, including those proposed in~\cite{Polegre2021} and~\cite{Bjoernson2020}.
Both of the bilinear-\ac{EP} algorithms execute 20 iterations for \ac{JCD}, whereas the \ac{VB-BP-EP} algorithm runs for 40 iterations.
All iterative algorithms use a damping parameter of $\eta=0.5$.
In the following figures, solid and dashed lines distinguish between different pilot sequences or data lengths, while colors and markers indicate the respective algorithms.

\begin{figure}[t]
    \centering
    \subfloat[$T_d=10$.]{\includegraphics[width=\plotwidth\textwidth]{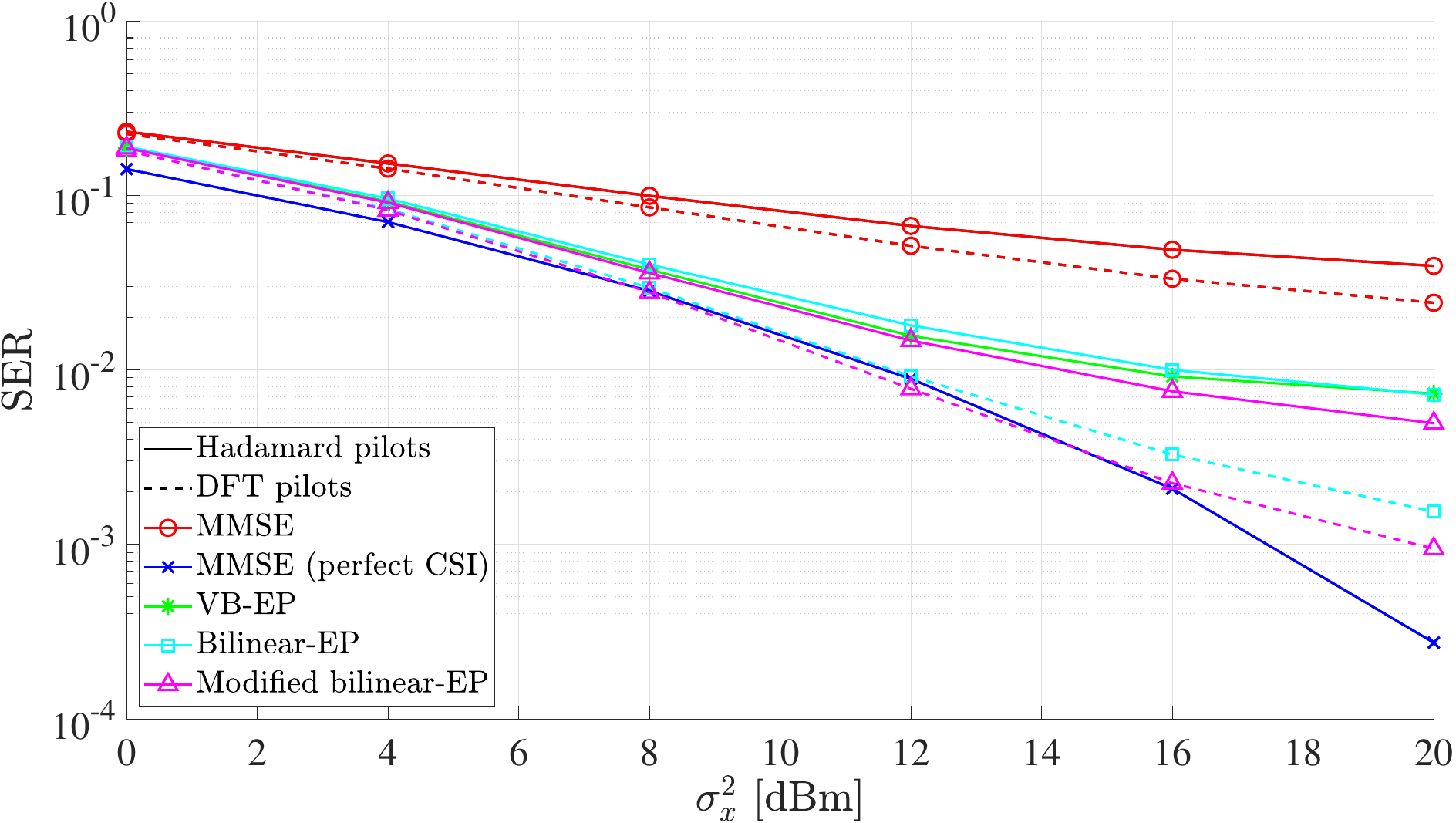}
    \label{fig:SER_SNR_Td10}}\vspace{-3mm}\\
    \subfloat[$T_d=30$.]{\includegraphics[width=\plotwidth\textwidth]{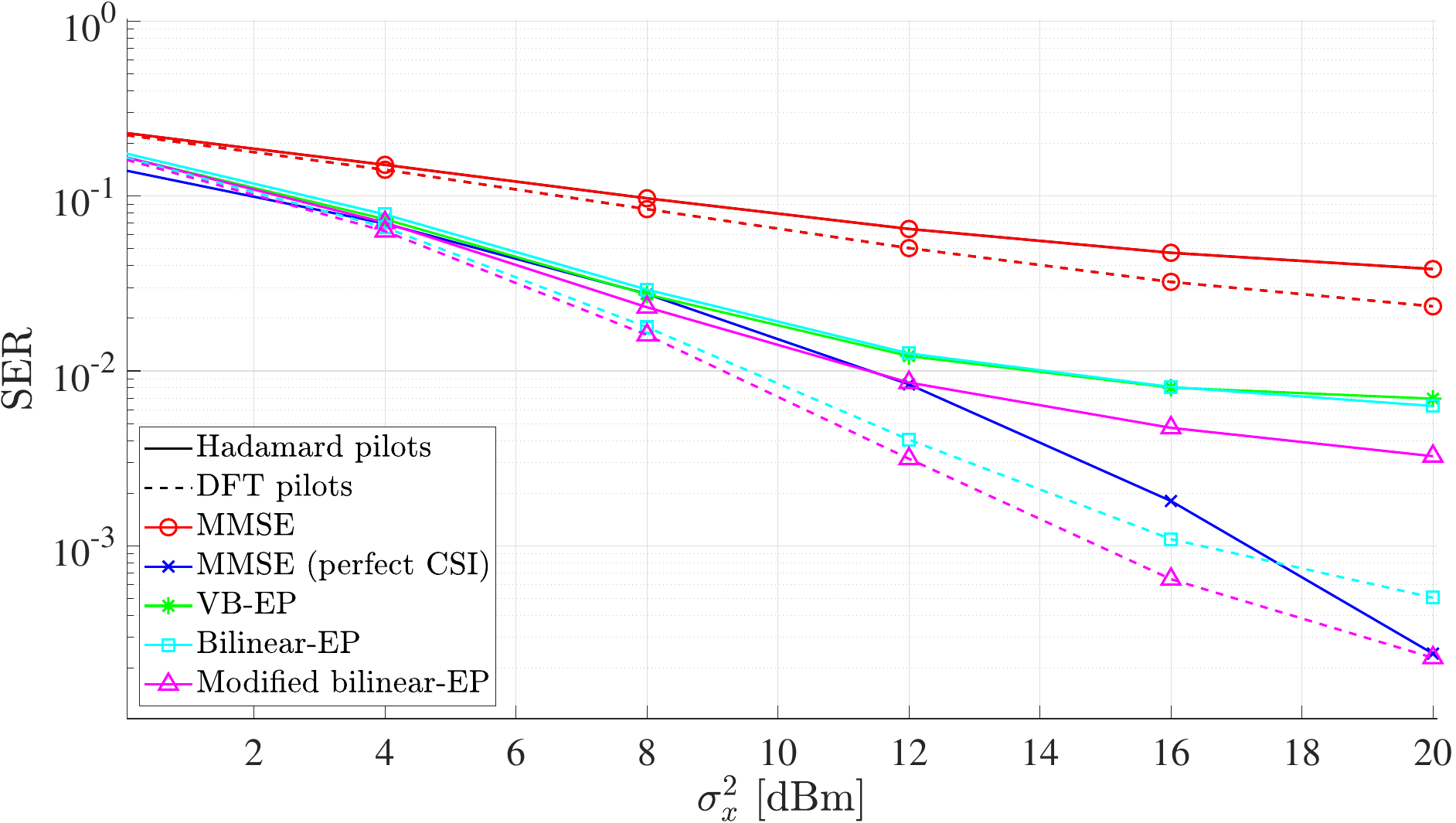}
    \label{fig:SER_SNR_Td30}}\\
    \caption{{\acs{SER}} versus transmit power.}
    \vspace*{-1mm}
    \label{fig:SER_SNR_Td10_Td30}
\end{figure}

\begin{figure}[t]
    \centerline{\includegraphics[width=\plotwidth\textwidth]{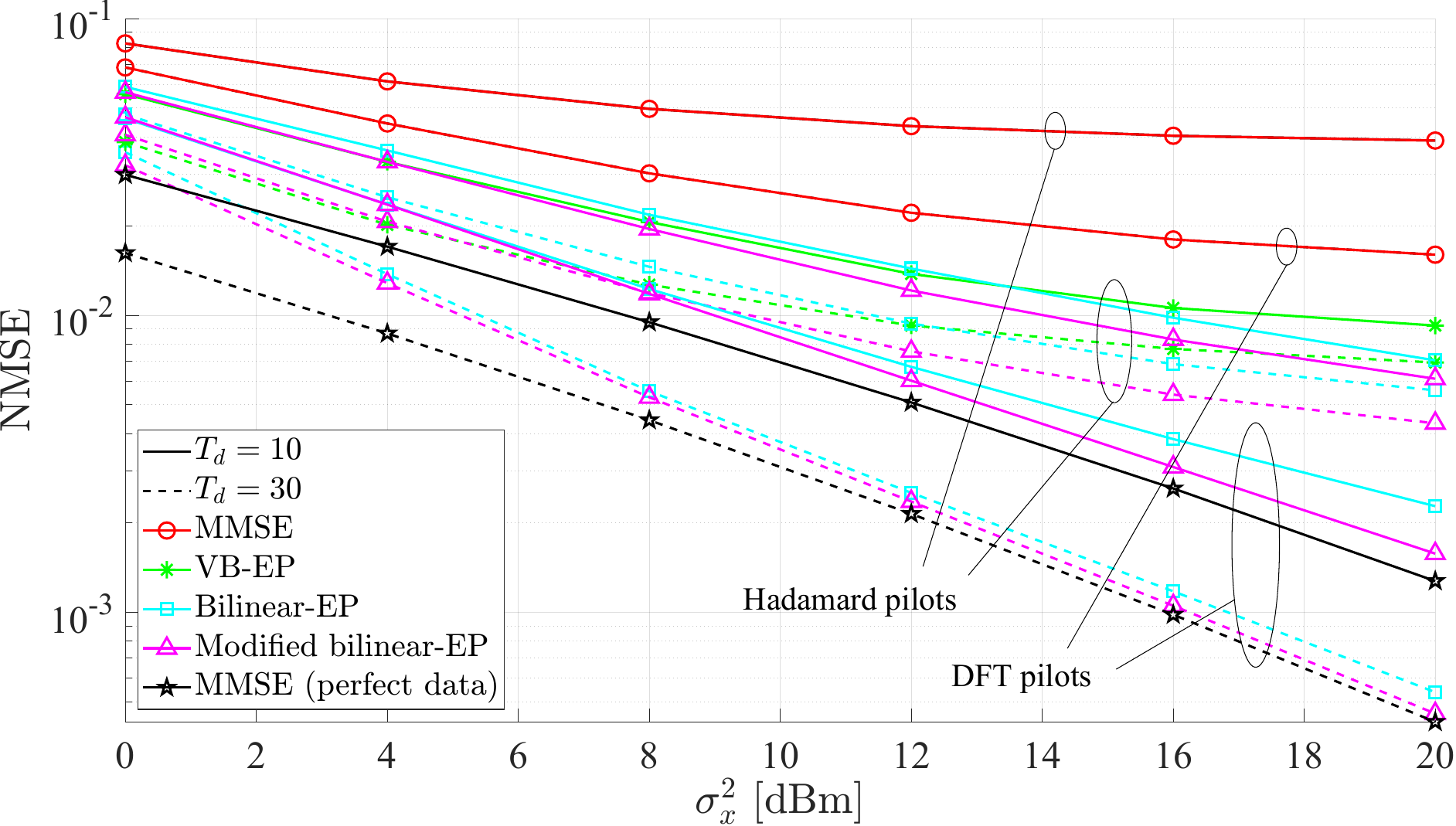}}
    \caption{\acs{NMSE} versus transmit power.}
    \vspace*{-1mm}
    \label{fig:NMSE_SNR_Td10_Td30}
\end{figure}

\begin{figure}[t]
    \centerline{\includegraphics[width=\plotwidth\textwidth]{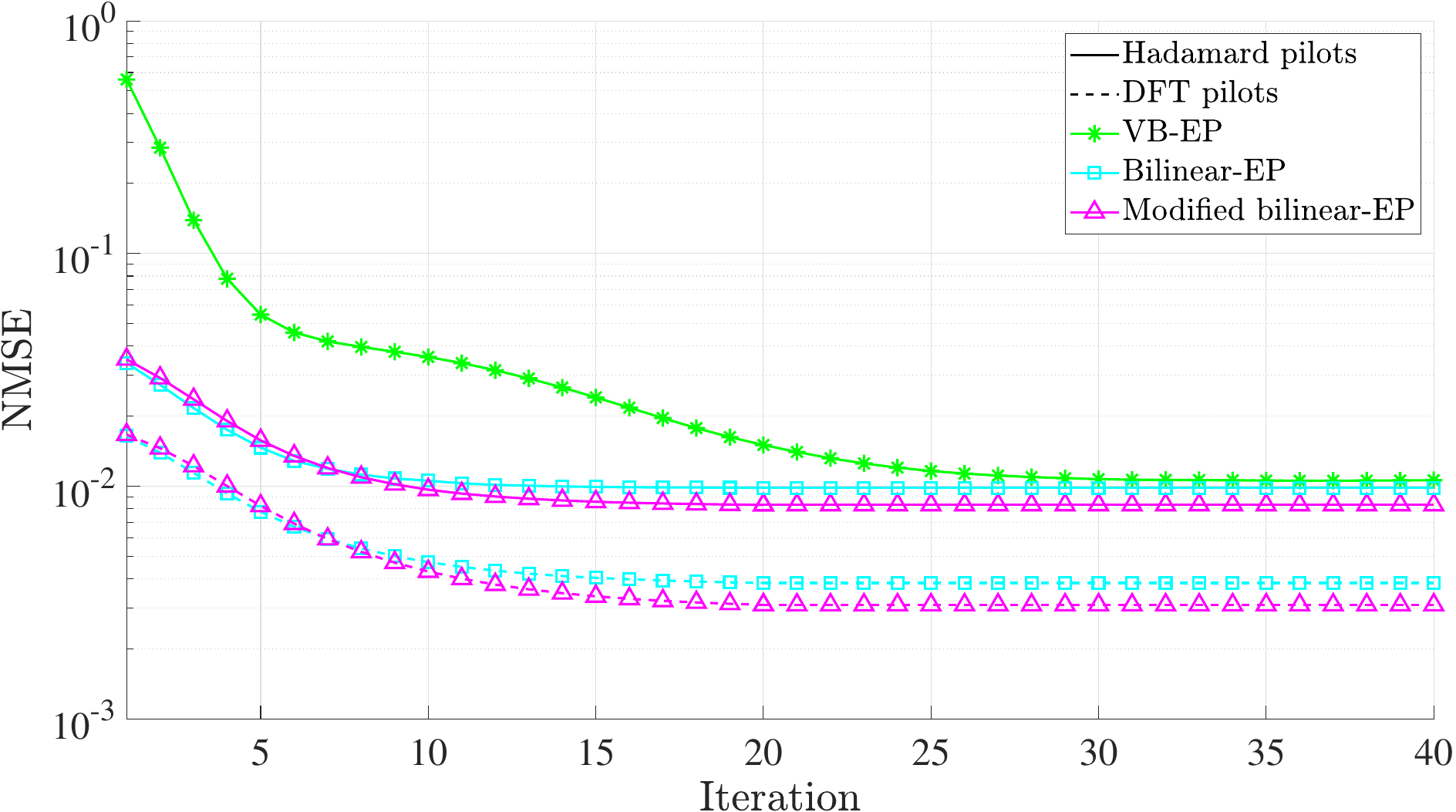}}
    \caption{\acs{NMSE} versus iterations for $T_d=10$ and $\sigma_x^2=16\,$dBm.}
    \vspace*{-1mm}
    \label{fig:NMSE_iter_Td10_16dBm}
\end{figure}

The next set of results present the \ac{NMSE} and the \ac{SER}  as functions of the \ac{UE} transmit power $\sigma_x^2$.
The \ac{NMSE} of the channel matrix estimate $\hat{\lmat{H}}$ is defined as $\text{NMSE}\coloneq\est{}{\frac{||\lmat{H}-\hat{\lmat{H}}||_F^2}{||\lmat{H}||_F^2}}$, and the \ac{SER} is obtained by averaging across all \acp{UE}, i.e., $\text{SER}\coloneq\est{}{\sum_k\sum_t\ind{\data{x}_{kt}\neq\data{\hat{x}}_{kt}} / (KT_d)}$.
The expectation operator in both definitions is computed with respect to the channel realizations.
In our simulation results, the performance is obtained by averaging $10^4$ block transmissions where each block transmission corresponds to an independent realization of the \ac{UE} positions.
The \ac{SER} results are shown in Fig.~\ref{fig:SER_SNR_Td10_Td30}.
It can be observed that the proposed bilinear-\ac{EP} algorithm outperforms the linear \ac{MMSE} detector as well as the bilinear-\ac{EP} algorithm in~\cite{Karataev2024} and the \ac{VB-BP-EP} algorithm in~\cite{Zhao2024}.
Systems adopting non-orthogonal \ac{DFT} pilots show significantly better performance compared to those using orthogonal Hadamard pilots.
Additionally, the performance gain from increasing the number of data symbols from $T_d=10$ to $T_d=30$ is more pronounced for \ac{DFT} pilots.
The \ac{VB-BP-EP} algorithm, being designed for orthogonal pilots, is not applicable to systems adopting the proposed \ac{DFT} pilots.
The proposed bilinear-\ac{EP} algorithm offers greater flexibility in the design of pilot sequences, which represents an additional advantage.
Fig.~\ref{fig:NMSE_SNR_Td10_Td30} illustrates the  \ac{NMSE} performance as a function of the transmit power  for the same settings discussed above.
The proposed bilinear-\ac{EP} algorithm, when used with non-orthogonal \ac{DFT} pilot sequences, closely approaches the performance of the genie-aided \ac{MMSE} estimator, a bound which is not attained with orthogonal Hadamard pilot sequences.
Fig.~\ref{fig:NMSE_iter_Td10_16dBm} shows the convergence behavior of the iterative algorithms.
The bilinear-\ac{EP} algorithm converges significantly  faster than the \ac{VB-BP-EP} algorithm.
As the bilinear-\ac{EP} algorithm is executed for only 20 iterations, the result for iteration number 20 is extended to all following iterations to enable direct comparison with the benchmark algorithms.

\begin{figure}[t]
    \centerline{\includegraphics[width=\plotwidth\textwidth]{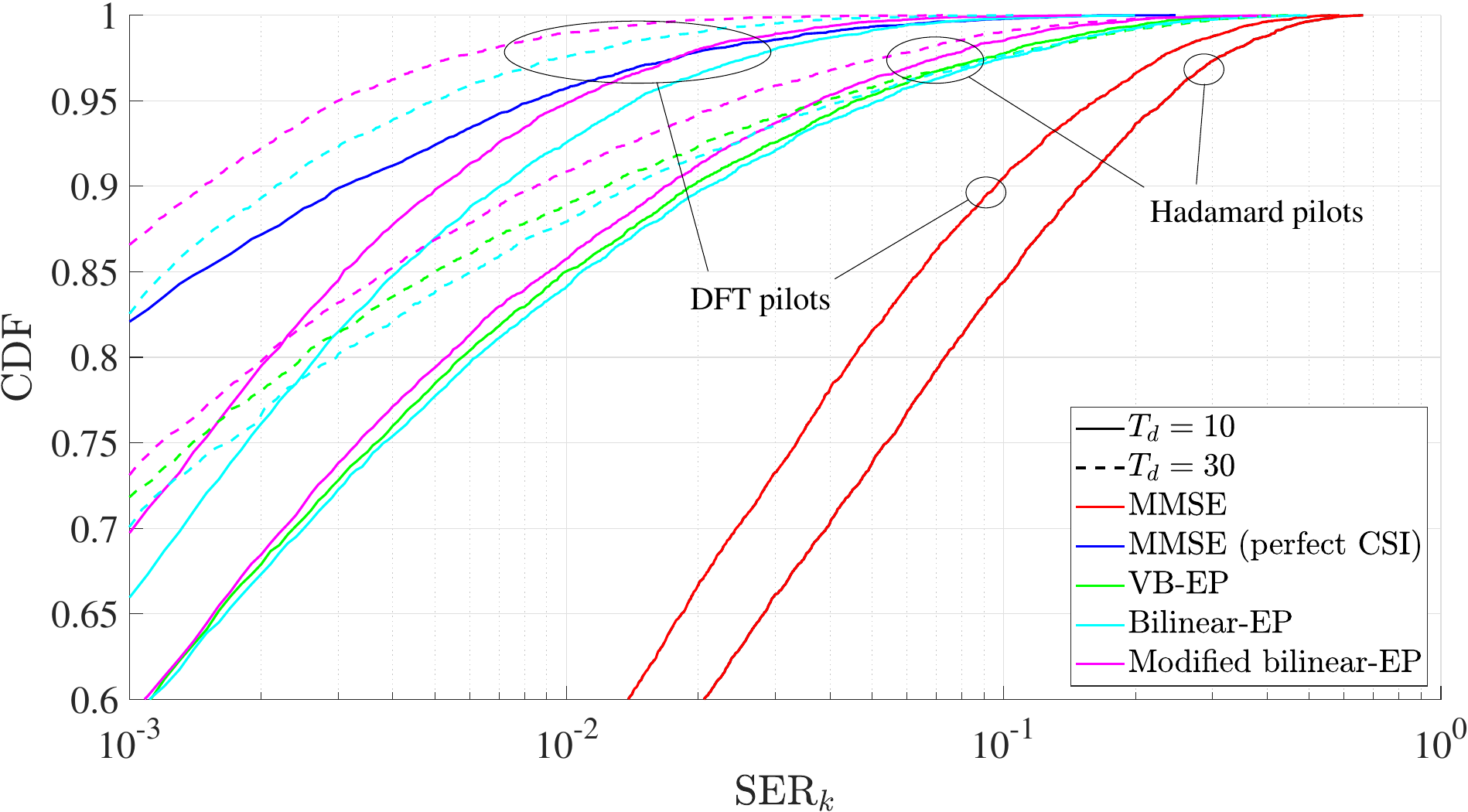}}
    \caption{\acs{CDF} of $\text{SER}_k$ for $\sigma_x^2=16\,$dBm.}
    \vspace*{-1mm}
    \label{fig:SER_CDF_Td10_Td30}
\end{figure}

\begin{figure}[t]
    \centerline{\includegraphics[width=\plotwidth\textwidth]{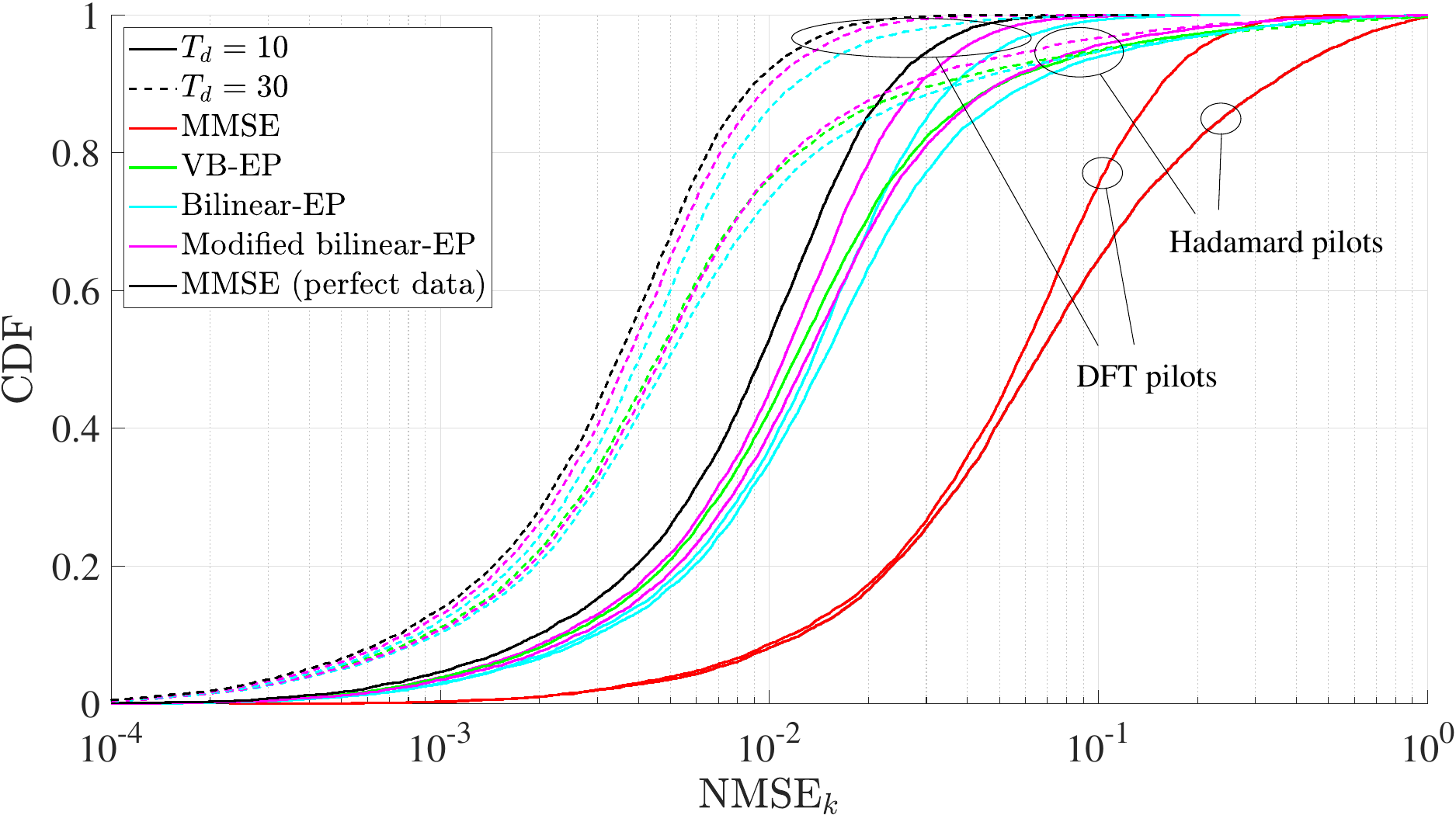}}
    \caption{\acs{CDF} of $\text{NMSE}_k$ for $\sigma_x^2=16\,$dBm.}
    \vspace*{-1mm}
    \label{fig:NMSE_CDF_Td10_Td30}
\end{figure}

For the following set of results, the \ac{UE} transmit power is fixed to $\sigma_x^2=16\,$dBm and the performance is assessed in terms of the \acp{CDF} of the \ac{NMSE} and the \ac{SER} per user, i.e., $\text{NMSE}_k\coloneq\est{}{\frac{||\lvec{h}_k-\hat{\lvec{h}}_k||^2}{||\lvec{h}_k||^2}}$ and $\text{SER}_k\coloneq\est{}{\sum_t\ind{\data{x}_{kt}\neq\data{\hat{x}}_{kt}} / T_d}$.
Here, 1000 independent realizations of the \ac{UE} positions are considered which results in $1000\cdot K=8000$ data points for the \ac{CDF}.
For each \ac{UE} positioning realization, the performance is averaged over 1000 independent block transmissions, accounting for small-scale fading and noise realizations.
The corresponding results are illustrated in Figs.~\ref{fig:SER_CDF_Td10_Td30} and~\ref{fig:NMSE_CDF_Td10_Td30}.
The \acp{CDF} show that the proposed modified bilinear-\ac{EP} algorithm outperforms the benchmark schemes.
Furthermore, the algorithms applied to systems utilizing non-orthogonal \ac{DFT} pilots show greater performance gains from increased data lengths, especially when considering the 95\%-likely performance.

\begin{figure}[t]
    \centerline{\includegraphics[width=\plotwidth\textwidth]{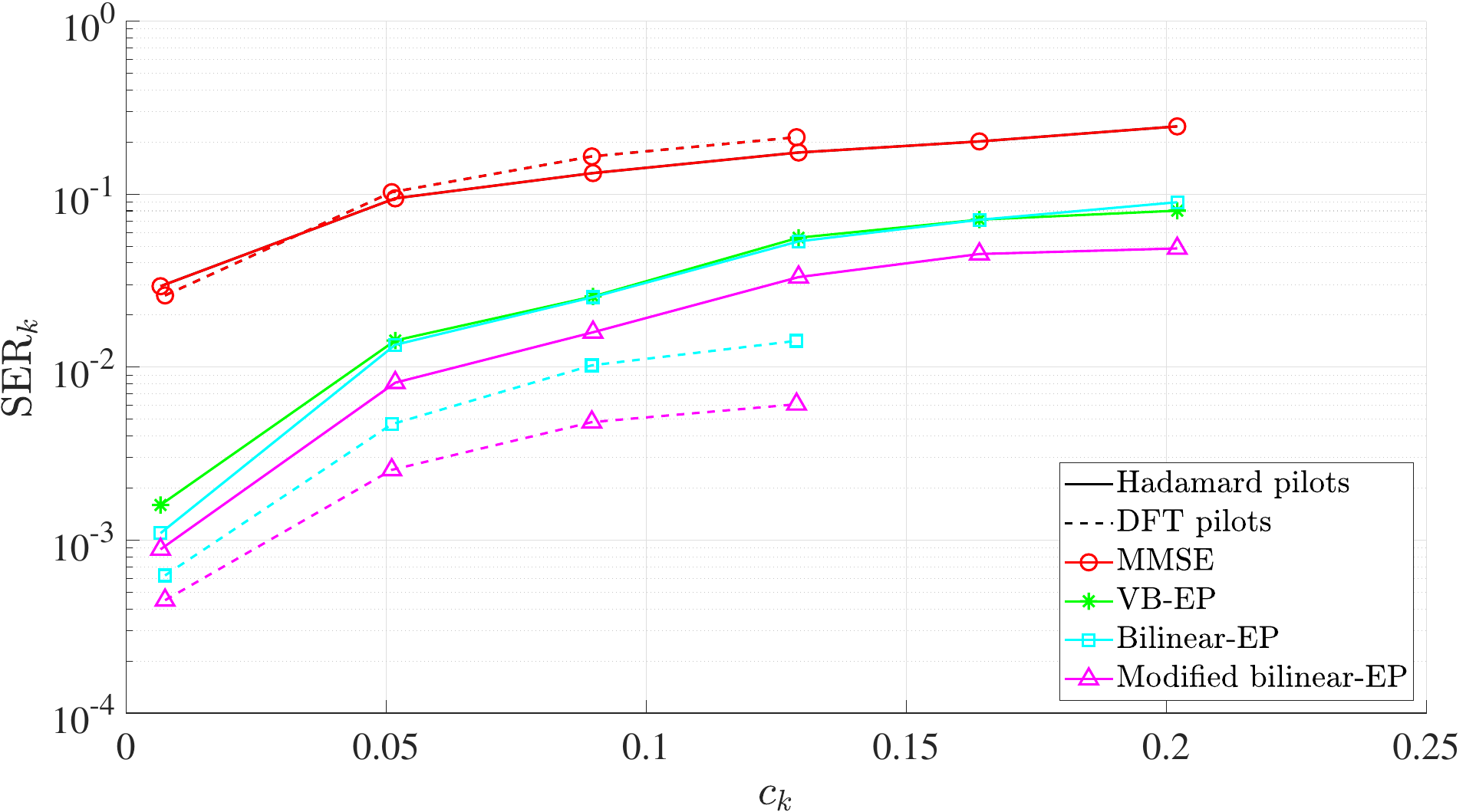}}
    \caption{$\text{SER}_k$ versus \acs{PC} metric for $T_d=30$ and $\sigma_x^2=16\,$dBm.}
    \vspace*{-1mm}
    \label{fig:SER_PC_Td30}
\end{figure}

For the final set of results, the exact same setup as before is used with the \ac{UE}-based performance metrics now evaluated  as a function of the \ac{PC} metric $c_k$.
The results are averaged over \acp{UE} experiencing a similar level of \ac{PC} quantified by $c_k$.
Fig.~\ref{fig:SER_PC_Td30} presents the corresponding results for the \ac{SER}.
It can be observed that as $c_k$ increases, the performance degrades which validates $c_k$ as an appropriate \ac{PC} metric.
Combined with the observations from Fig.~\ref{fig:c_k_CDF}, it explains the superior average performance of \ac{DFT} pilots over Hadamard pilots in previous results.
Furthermore, the proposed algorithm consistently achieves the  best performance  for a given level of \ac{PC}.
These results further show that  the \ac{JCD} schemes applied to systems using \ac{DFT} pilots are more effective at mitigating \ac{PC} for a given level of \ac{PC} $c_k$.

\section{Conclusion}\label{sec:concl}
In this work, we proposed a novel \ac{JCD} algorithm based on \ac{EP}, designed to improve robustness against \ac{PC} in \ac{CF-MaMIMO} systems.
The algorithm extended the bilinear-\ac{EP} method in~\cite{Karataev2024} and achieved improved channel estimation and data detection performance, particularly under severe \ac{PC} and for systems using non-orthogonal pilot sequences.
It consistently outperformed optimal linear detectors in~\cite{Polegre2021} and~\cite{Bjoernson2020} and state-of-the-art \ac{JCD} algorithms.
We also compared systems employing orthogonal and non-orthogonal pilots and showed that non-orthogonal sequences provided significant performance gains.
Finally, we introduced a new metric to quantify the impact of \ac{PC} on iterative \ac{JCD} algorithms and demonstrated its relevance and consistency.

\bibliographystyle{IEEEtran}
\bibliography{IEEEabrv,references_abbrev}

\end{document}